\documentclass[sigplan,screen]{acmart}

\setcopyright{acmlicensed}
\copyrightyear{2026}
\acmYear{2026}
\acmConference[ASPLOS '26]{International Conference on Architectural Support for
Programming Languages and Operating Systems}{March 22--26, 2026}{Pittsburgh, PA}

\usepackage{format/custom_package}
\usepackage{format/macro}

\begin{document}

\title{Neo: Real-Time On-Device 3D Gaussian Splatting with Reuse-and-Update Sorting Acceleration}

\author{Changhun Oh}
\orcid{https://orcid.org/0009-0009-9834-3848}
\affiliation{
  \institution{KAIST}
  \city{Daejeon}
  \country{Republic of Korea}
}
\email{choh@casys.kaist.ac.kr}

\author{Seongryong Oh}
\orcid{https://orcid.org/0009-0004-6707-0641}
\affiliation{
  \institution{KAIST}
  \city{Daejeon}
  \country{Republic of Korea}
}
\email{sroh@casys.kaist.ac.kr}

\author{Jinwoo Hwang}
\orcid{https://orcid.org/0009-0008-8498-2502}
\affiliation{
  \institution{KAIST}
  \city{Daejeon}
  \country{Republic of Korea}
}
\email{jwhwang@casys.kaist.ac.kr}

\author{Yoonsung Kim}
\orcid{https://orcid.org/0009-0000-2333-292X}
\affiliation{
  \institution{KAIST}
  \city{Daejeon}
  \country{Republic of Korea}
}
\email{yskim@casys.kaist.ac.kr}

\author{Hardik Sharma}
\orcid{https://orcid.org/0000-0003-0028-013X}
\affiliation{
  \institution{Meta}
  \city{Sunnyvale}
  \state{CA}
  \country{USA}
}
\email{hardiksharma@meta.com}

\author{Jongse Park}
\orcid{https://orcid.org/0000-0002-6629-449X}
\affiliation{
  \institution{KAIST}
  \city{Daejeon}
  \country{Republic of Korea}
}
\email{jspark@casys.kaist.ac.kr}

\begin{abstract}

3D Gaussian Splatting (3DGS) rendering in real-time on resource-constrained devices is essential for delivering immersive augmented and virtual reality (AR/VR) experiences.
However, existing solutions struggle to achieve high frame rates, especially for high-resolution rendering.
Our analysis identifies the sorting stage in the 3DGS rendering pipeline as the major bottleneck due to its high memory bandwidth demand.
This paper presents~\sysname{}, which introduces a reuse-and-update sorting algorithm that exploits temporal redundancy in Gaussian ordering across consecutive frames and devises a hardware accelerator optimized for this algorithm.
By efficiently tracking and updating Gaussian depth ordering instead of re-sorting from scratch,~\sysname{} significantly reduces redundant computations and memory bandwidth pressure.
Experimental results show that~\sysname{} achieves up to 10.0$\times$ and 5.6$\times$ higher throughput than state-of-the-art edge GPU and ASIC solution, respectively, while reducing DRAM traffic by 94.5\% and 81.3\%.
These improvements make high-quality and low-latency on-device 3D rendering more practical.

\end{abstract}



\maketitle 

\section{Introduction}
\label{sec:introduction}

The success of Generative AI in text generation~\cite{gpt-4, deepseek-v2, chat-gpt, OPT, llama, claude, guo2025deepseek, borgeaud2022improving, lewis2020retrieval, dao2022flashattention, kwon2023efficient} has demonstrated its potential, paving the way for researchers to explore its next step: virtual world generation~\cite{bautista2022gaudi, qian2024magic123, kim2025mixdit, kong2024cambricon, liu2024sora, esser2024scaling, ho2020denoising, rombach2022high, ho2022video, blattmann2023stable}.
Modern augmented and virtual reality (AR/VR) devices such as Apple Vision Pro~\cite{apple-vision-pro-tech-spec}, Meta Orion AR Glasses~\cite{meta-orion}, and Meta Quest 3~\cite{meta-quest-3} exemplify the growing demand for immersive environments, where human interactions unfold in dynamic, computer-generated worlds, necessitating high-fidelity content and scalable virtual world creation.

\emph{View synthesis}, which creates new viewpoints of a scene from existing data, is crucial for enabling virtual experiences.
Recently, 3D Gaussian Splatting (3DGS) has emerged as an effective technique that balances rendering quality and performance, while providing a low-cost solution for immersive applications.
To ensure a smooth user experience~\cite{wang2023effect, geris2024balancing}, such applications demand high-resolution rendering with ultra-low latency (e.g., 7–15ms\cite{ars-technica:2013:effective-latency-of-vr}).
While cloud processing may appear attractive, it introduces critical delays~\cite{bhuyan2024gamestreamsr, bhuyan2022end}, thereby necessitating on-device 3DGS rendering.

However, this on-device requirement creates tension with the high computational demands of real-time view synthesis.
For instance, industry-leading automotive GPUs such as the Jetson Orin~\cite{nvidia-orin-series,karumbunathan2022nvidia} and even the state-of-the-art 3DGS-optimized accelerator~\cite{lee2024gscore} deliver only around 60 FPS at HD resolution, offering lower frame rates when targeting the per-eye high-resolution commonly used in AR/VR, such as 2K or 4K per eye~\cite{apple-vision-pro-tech-spec,sie-playstation-vr2,hololens2,pimax-crystal, hp-reverb-g2, varjo-aero}.
This performance gap underscores the need for more efficient on-device solutions.

\begin{figure}[t]
        \vspace{-\baselineskip}
        \centering
        \includegraphics[width=\linewidth]{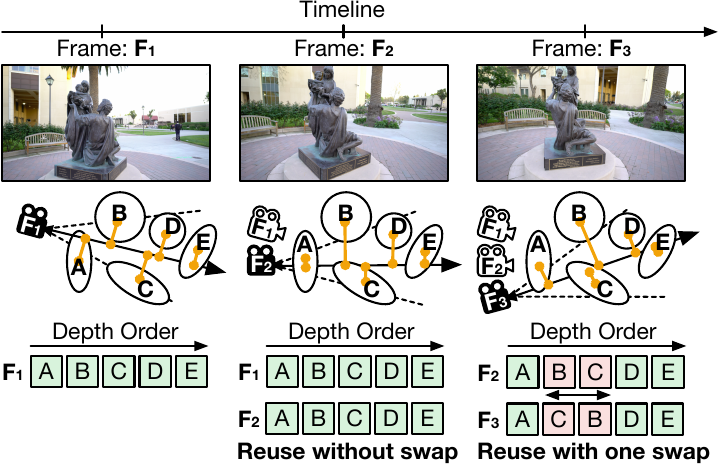}
        \caption{Reuse opportunities in the sorting stage of 3D Gaussian Splatting (3DGS) inference. The figure illustrates how the Gaussian order across three consecutive frames (F1, F2, and F3) exhibits significant temporal similarities.
        }
        \label{fig:introduction}
\end{figure}

To better understand the bottlenecks and opportunities, we first conduct a thorough performance characterization of an existing 3DGS accelerator, GSCore~\cite{lee2024gscore}. 
Our analyses suggest that GSCore effectively mitigates bottleneck in rasterization, leaving the sorting stage as a new and dominant bottleneck in the rendering pipeline.
Unfortunately, sorting is a well-established operation in modern computing, making it inherently difficult to accelerate through conventional means. 
However, 3DGS renders frames sequentially over time, repeatedly performing sorting on largely similar Gaussian orders across consecutive frames.
This temporal redundancy presents an opportunity to minimize redundant computations and improve efficiency.
Inspired by this insight, this paper proposes~\sysname{} by jointly designing (1) reuse-and-update sorting mechanism to exploit temporal redundancy in Gaussian ordering and (2) a hardware-accelerated sorting pipeline for on-device 3DGS rendering.
Figure~\ref{fig:introduction} depicts the reuse opportunities that form the basis of the proposed reuse-and-update sorting mechanism.
In designing~\sysname, we identify the following challenges:
\begin{itemize}[labelindent=0.3em,nolistsep,leftmargin=1.0em]
\item \textbf{Challenge 1: High sorting overhead in rendering.}
In general, sorting is performed occasionally and reused across iterations.
However, sorting in 3DGS differs from general purpose sorting, as it requires reordering millions of Gaussians for every rendered frame.
Existing sorting implementations rely on bandwidth-intensive memory accesses, making real-time on-device rendering impractical.
\item \textbf{Challenge 2: Missed temporal redundancy in sorting.}
Re-sorting Gaussians from scratch every frame overlooks the strong temporal locality inherent in 3DGS rendering. Since most Gaussians retain similar depth order across consecutive frames, treating each frame independently leads to redundant computations and excessive DRAM traffic, limiting scalability at high resolutions.
\end{itemize}

To address the aforementioned challenges, this paper makes the following contributions.

\begin{itemize}[labelindent=0.3em,nolistsep,leftmargin=1.0em]

\item \textbf{Reuse-and-update sorting for temporal redundancy exploitation.}  
~\sysname{} introduces a reuse-and-update sorting mechanism that leverages the temporal redundancy in 3D Gaussian Splatting (3DGS) rendering.  
Instead of sorting Gaussians from scratch for every frame,~\sysname{} efficiently tracks and updates the sorted Gaussian table across consecutive frames.
By identifying minimal changes in Gaussian ordering over time, this mechanism significantly reduces redundant sorting computations and memory bandwidth overhead.  
Moreover, by selectively updating only designated segments of the table,~\sysname{} minimizes unnecessary operations, thereby improving overall processing efficiency.
As a result, this optimization enables real-time sorting at AR/VR resolutions while maintaining rendering accuracy, even in scenes with frequent viewpoint changes.

\item \textbf{Accelerating sorting stage for on-device rendering.}  
\sysname{} implements a hardware-accelerated sorting pipeline optimized for efficient on-device 3DGS rendering.
While existing accelerators focus primarily on optimizing rasterization, they overlook the inefficiencies of sorting, resulting in excessive DRAM traffic.
~\sysname{} addresses this limitation by integrating dedicated sorting hardware that minimizes memory access overhead and accelerates depth-based Gaussian ordering within the rendering pipeline.
This hardware adopts a hybrid sorting strategy that integrates both partial and global ordering, effectively performing Gaussian sorting with low computational overhead.
%
%
By coupling this hardware support with a reuse-and-update sorting mechanism, ~\sysname{} delivers significant improvements in both latency and throughput, making real-time, high-resolution AR/VR rendering practical.
\end{itemize}

To evaluate~\sysname{}, we implement a cycle-accurate simulator based on the 3DGS rendering pipeline and model the execution of sorting operation.
Our evaluation uses a representative set of neural rendering workloads, including complex scenes to assess the impact of temporal redundancy. 
We synthesize~\sysname{} using Synopsys Design Compiler with ASAP 7nm library~\cite{asap7} and measure its performance and memory bandwidth usage against NVIDIA Orin AGX GPU~\cite{karumbunathan2022nvidia} and the state-of-the-art accelerator GSCore~\cite{lee2024gscore}.
Our results show that~\sysname{} achieves up to 10.0$\times$ and 5.6$\times$ higher throughput compared to the Orin AGX GPU and GSCore, respectively, while reducing sorting-induced memory traffic by 94.4\% and 81.3\%.
Furthermore,~\sysname{} enables real-time 3DGS rendering at QHD resolutions, achieving the an average throughput of 99.3 FPS required for smooth AR/VR experiences.
These results demonstrate that~\sysname{} effectively overcomes the sorting bottleneck in real-time 3DGS rendering, advancing the realization of generative virtual worlds and enabling truly immersive on-device experiences.
\begin{figure*}[t]
    \centering
    \includegraphics[width=\linewidth]{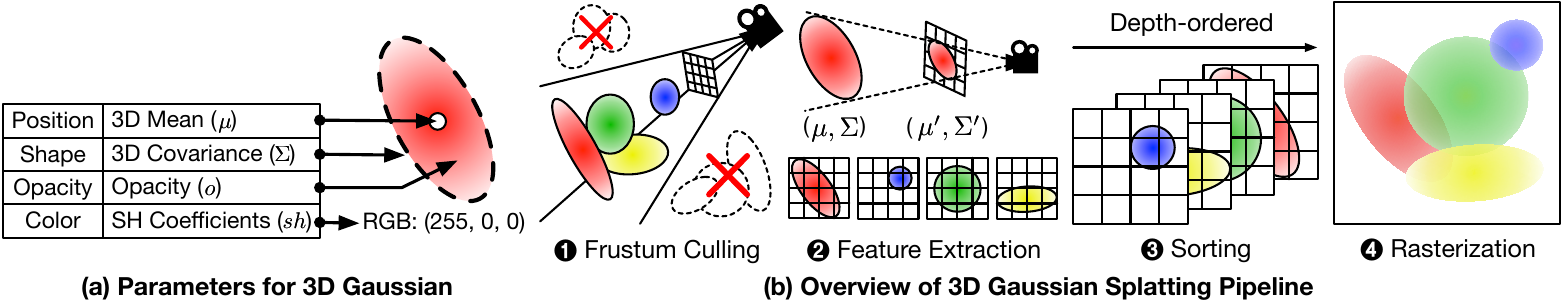}
    \caption{Brief overview of 3D Gaussian Splatting.}
    \label{fig:background-3dgs-overview}
    \vspace{-\baselineskip}
\end{figure*}

\section{Background}

\subsection{Real-Time 3D Rendering in On-Device Systems}
\label{subsec:3d-rendering-in-on-device-systems}

Recent advances in augmented and virtual reality (AR/VR) have driven the development of mobile devices and headsets \cite{apple-vision-pro-tech-spec, meta-quest-3, htc-vive-xr-elite, pico-4-ultra, sie-playstation-vr2} capable of rendering high-fidelity scenes. 
These on-device platforms support high frame rates and resolution, delivering immersive experiences.
For instance, Apple Vision Pro~\cite{apple-vision-pro-tech-spec} features a combined 23 million pixels, achieving 4K-level resolution.
Meanwhile, Meta Quest 3~\cite{meta-quest-3} provides a per-eye resolution of 2064$\times$2208, with both devices supporting refresh rates of up to 90Hz. 
These stringent specifications are critical for delivering immersive experiences and ensuring user comfort~\cite{wang2022effect, wang2023effect}.
However, meeting these demands is challenging when rendering high-fidelity scenes given the limited computing resources of on-device systems. 
A straightforward approach is offloading rendering requests to cloud servers. However, this method suffers from frame drops due to network congestion, which is a critical issue in latency-sensitive rendering applications~\cite{bhuyan2024gamestreamsr, bhuyan2022end}.
These constraints highlight the necessity for on-device AR/VR platforms to integrate on-device rendering capabilities.

\subsection{3D Gaussian Splatting (3DGS)}

3DGS~\cite{kerbl20233d} has emerged as a promising rendering method for synthesizing complex real-world scenes while achieving real-time rendering performance. 
To render 3D scenes, 3DGS exploits millions of 3D Gaussians, modeled as anisotropic ellipsoids. 
As illustrated in Figure~\ref{fig:background-3dgs-overview}(a), each Gaussian is identified by a radial opacity $\alpha$, as shown in Equation~\ref{eqn:3D-gaussian-definition}:
\setlength{\abovedisplayskip}{1ex}
\setlength{\belowdisplayskip}{1ex}
\begin{align}
\alpha (x) = o \cdot e^{-\frac{1}{2}(x - \mu)^T\mathbf{\Sigma}^{-1}(x - \mu)}
\label{eqn:3D-gaussian-definition}
\end{align}
where $o$ is an opacity value, $\boldsymbol{\mu}$ is a mean vector, $\boldsymbol{\Sigma}$ is a 3D covariance matrix. 
In addition, $sh$ refers to spherical harmonics coefficients, which enable rendering of view-dependent color~\cite{green2003spherical}. 
These explicit 3D representations are learnable parameters, trained with differentiable rasterization and gradient-based optimization.
%
Given a set of Gaussians, 3DGS employs a $\alpha$-blending process~\cite{porter1984compositing}, akin to traditional rendering methods~\cite{brebin1998volume} that use explicit 3D representations to render scenes. 
This innovative approach incorporates the strengths of both traditional scene reconstruction methods~\cite{levoy2023light, waechter2014let, wood2023surface} and prior neural rendering techniques~\cite{mildenhall2021nerf, garbin2021fastnerf, muller2022instant, reiser2021kilonerf}, delivering exceptional rendering quality and performance.

\subsection{3D Gaussian Splatting Pipeline}
\label{subsec:3dgs-pipeline}

Figure~\ref{fig:background-3dgs-overview}(b) denotes the 3DGS pipeline with four main stages: frustum culling, feature extraction, sorting, and rasterization. 

\niparagraph{\circleN{1} Frustum Culling.}
First, the system discards Gaussians that are not visible from the current camera viewpoint and preserves only those required for subsequent stages. 
This initial filtering process reduces redundant computations for Gaussians outside the camera's field of view. 

\niparagraph{\circleN{2} Feature Extraction.} 
With the filtered 3D Gaussians, the system projects them onto the camera's image plane, extracting their view-dependent features in conjunction with the camera viewpoint. This includes 2D representations $(\mu^\prime, \boldsymbol{\Sigma}^\prime)$ transformed by 3D parameters $(\mu, \boldsymbol{\Sigma})$ and color ($c$) calculated using the coefficients of spherical harmonics $(sh)$~\cite{green2003spherical}. 

\niparagraph{\circleN{3} Sorting.} 
After extracting features from 3D Gaussians, the projected 2D Gaussians overlap on the image plane. 
At this stage, the system sorts them by depth. 
This ordering is a crucial step for the subsequent rasterization stage, which leverages it to blend overlapping 2D Gaussians by accumulating their pixel colors in a depth-sorted manner. 

\niparagraph{\circleN{4} Rasterization.} 
At this stage, the system computes pixel colors by $\alpha$-blending~\cite{porter1984compositing} the depth-sorted Gaussians.
Starting from the foremost Gaussian, each Gaussian contributes to the pixel color and accumulates opacity.
When the cumulative opacity exceeds a predefined threshold, further processing for that pixel stops, as its color is considered finalized, thereby reducing unnecessary computation.

\subsection{3D Gaussian Splatting Acceleration}
\niparagraph{Tile-based parallelism.}
For efficient rendering, the 3DGS system subdivides the image plane into a grid of smaller 2D regions (\emph{tiles}).
In the \circleN{3} sorting stage, it duplicates and distributes Gaussians to the intersected tiles, and in the \circleN{4} rasterization stage, it processes 2D Gaussians on a \emph{per-tile} basis.
This approach enables the system to process only the Gaussians within each tile's boundary, thereby reducing redundant computation.
Furthermore, it leverages hardware threads to process multiple tiles in parallel, further improving overall rendering performance.

\niparagraph{GPU implementations.}
To leverage this parallelism, 3DGS methods~\cite{kerbl20233d, compgs:eccv:24, papantonakis:i3d:24} employ GPUs for tile-based sorting and rasterization. 
They utilize NVIDIA CUB library~\cite{nvidia-cccl} for sorting and implement custom CUDA kernels~\cite{kopanas:2021:point-based-neural-rendering} to support a cumulative $\alpha$-blending during rasterization.
Although these GPU-driven approaches enhance rendering performance through extensive parallel processing, the cumulative $\alpha$-blending step remains a significant bottleneck, especially in on-device AR/VR environments.

\niparagraph{ASIC acceleration: GSCore.}
To address this challenge, recent work GSCore~\cite{lee2024gscore} introduces a pioneering ASIC-based acceleration solution for on-device 3DGS systems. 
By adopting hierarchical tile-based sorting and subtile-based rasterization, GSCore efficiently processes millions of Gaussians under tight resource constraints. 
While this specialized co-design significantly outperforms GPU-based solutions, the stringent on-device constraints continue to push resource demands beyond what GSCore approaches can readily handle.
In the following section, we analyze how these constraints affect on-device 3DGS performance and define the key research challenge of this work.
\section{Motivation}

\begin{figure}[t]
    \centering
    \includegraphics[width=\linewidth]{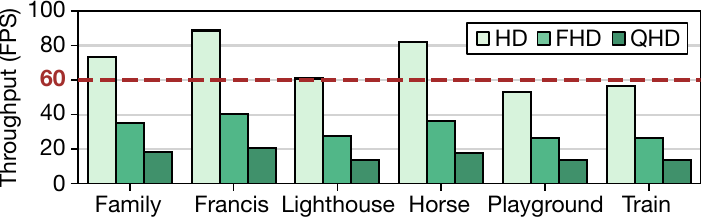}
    \caption{Throughput comparison with different resolutions.}
    \label{fig:motivation-gscore-varying-resolution}
    \vspace{-\baselineskip}
\end{figure}

\begin{figure}[t]
    \centering    \includegraphics[width=\linewidth]{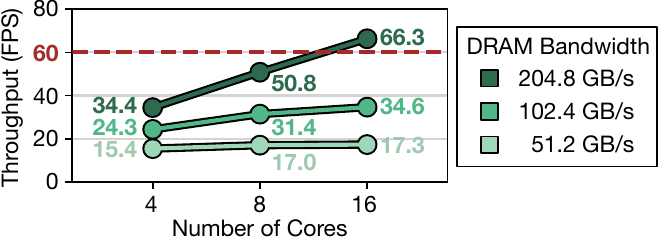}
    \caption{Throughput comparison across core counts and DRAM bandwidth when rendering at QHD resolution. Each colored label denotes the corresponding FPS performance.}
    \label{fig:motivation-gscore-varying-core-bandwidth}
    \vspace{-\baselineskip}
\end{figure}

\subsection{Challenge in On-Device 3DGS Rendering}

Despite GSCore’s efforts to accelerate 3DGS rendering, achieving real-time performance under the high frame rate and resolution requirements of modern AR/VR platforms (see Section~\ref{subsec:3d-rendering-in-on-device-systems}) remains a significant challenge.
To analyze these characteristics, we selected six scenes from the Tanks and Temples dataset~\cite{tankandtemples:tog:2017}: Family, Francis, Horse, Lighthouse, Playground, and Train, which were captured in real-world outdoor environments and serve as representative benchmarks for evaluating the performance of 3DGS rendering.

\niparagraph{Rendering performance under constraints.}
Figure~\ref{fig:motivation-gscore-varying-resolution} shows the throughput (FPS) performance of GSCore at three different resolutions: HD (1280$\times$720), FHD (1920$\times$1080), and QHD (2560$\times$1440). 
Following the rationale described in the original paper~\cite{lee2024gscore}, we configure the evaluation system with 4 computing cores and a DRAM bandwidth of 51.2 GB/s. 
This configuration enables a thorough assessment of GSCore's FPS performance within the tight resource budgets typical of on-device AR/VR environments. 
The results show that GSCore achieves 66.7~FPS at HD (1280$\times$720), exceeding the conservative service-level objective (SLO) requirement of 60~FPS. 
However, at higher resolutions GSCore experiences a significant FPS drop, achieving only 31.1~FPS and 15.8~FPS at FHD and QHD, respectively. 
These performance limitations underscore the need for a robust acceleration solution capable of delivering scalable and responsive 3DGS rendering within resource-constrained on-device systems. 

\subsection{Performance Characterization of 3DGS}
\label{subsec:bottleneck-analysis}
To better understand the throughput performance of 3DGS at high-resolution (QHD), we perform a bottleneck analysis of GSCore by varying two primary system knobs: the number of compute unit cores and available DRAM bandwidth. 
Figure~\ref{fig:motivation-gscore-varying-core-bandwidth} shows FPS performance against varying core counts (4, 8, and 16 cores) under three distinct DRAM bandwidth conditions (51.2 GB/s, 102.4 GB/s, and 204.8 GB/s). 

\niparagraph{Implication from cores.}
First, we focus on the scenario representative of typical edge devices, characterized by limited DRAM bandwidth around 51.2 GB/s, as provided by conventional GSCore setups.
Under this bandwidth constraint, increasing the core count from 4 to 16 provides minimal performance improvements; specifically, even a fourfold increase in cores yields only about a 1.12$\times$ improvement in FPS, falling significantly short of the targeted 60 FPS SLO. 
This result indicates that computational scaling alone is insufficient when bandwidth constraints are tight.

\begin{figure}[t]
    \centering
    \includegraphics[width=\linewidth]{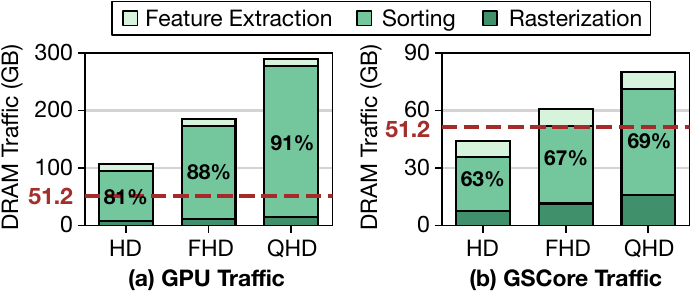}
    \caption{DRAM traffic (GB) required for rendering 60 frames and breakdown of memory bandwidth consumption across 3DGS pipeline stages.}
    \label{fig:motivation-traffic}
    \vspace{-\baselineskip}
\end{figure}

\niparagraph{Implication from bandwidth.}
Next, we examine the impact of DRAM bandwidth on the FPS performance of the 3DGS system. 
At the highest bandwidth (204.8~GB/s), which is a 4$\times$ increase over a typical on-device system (51.2~GB/s), the 3DGS system surpasses the 60~FPS SLO, achieving a 3.83$\times$ improvement in FPS. 
These results indicate that high-resolution 3DGS performance is constrained by DRAM bandwidth rather than computational power, identifying memory bandwidth as the primary bottleneck.

\niparagraph{Bandwidth breakdown.}
To further investigate the sources of bandwidth usage, we perform an analysis of bandwidth consumption in high-resolution 3DGS. 
Figure~\ref{fig:motivation-traffic} presents the DRAM traffic (GB) required to render 60 frames and its breakdown for the GPU-based 3DGS and GSCore across different resolutions.
In both systems, the results show that the sorting stage dominates bandwidth consumption, accounting for up to 90.8\% on the GPU and 69.3\% on GSCore.
GSCore achieves a noticeable reduction in memory requirements for 3DGS rendering.
However, it still requires an average bandwidth of 60.7~GB/s and 80.0~GB/s in FHD and QHD resolutions, respectively. 
This high bandwidth demand limits the availability of high-resolution 3DGS rendering in on-device AR/VR systems, whose practical DRAM bandwidths range from 17.8 GB/s to 59.7 GB/s~\cite{hill2019gables, lee2024gscore,li2022rt,gennerf:isca:23,neurex:isca:23,instant3d:isca:23,ngpc:isca:23, li2025uni}. 
These findings highlight the need to optimize the sorting stage, which is the major bandwidth bottleneck in the 3DGS pipeline, to realize high-resolution and low-latency on-device 3DGS systems. 

\subsection{Opportunity to Reduce Computation in Sorting}
\label{sec:opportunity-temporal-similarity}

To alleviate the bandwidth bottleneck caused by Gaussian sorting, we investigate the potential of temporally reusing Gaussians previously sorted in prior frames. 

\niparagraph{Per-frame sorting in 3DGS.}
In Gaussian sorting, each tile identifies the Gaussians that intersect it, determines their depth order for the subsequent rasterization stage, and stores this order in a Gaussian table.
As the camera viewpoint changes, this table becomes outdated, so the system reprocesses the sorting stage from scratch for every frame.
However, per-frame sorting overlooks the temporal similarity between consecutive frames, which arises from the gradual motion of Gaussians during camera movement and results in unnecessary memory bandwidth consumption.

\begin{figure}[t]
    \centering
    \includegraphics[width=\linewidth]{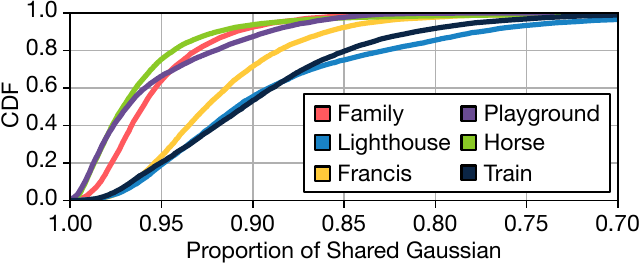}
    \caption{Temporal similarity of assigned Gaussian per tile.}
    \label{fig:opportunity-similarity-assigned-gaussian}
    \vspace{-\baselineskip}
\end{figure}

\niparagraph{Temporal similarity analysis.}
To quantify this temporal similarity, we first analyze the Gaussian retention between the Gaussian tables of consecutive frames.
Figure~\ref{fig:opportunity-similarity-assigned-gaussian} shows the cumulative distribution function (CDF) of the proportion of shared Gaussians across six real-world scenes.
In all scenes, over 90\% of tiles retain more than 78\% of their Gaussians from the previous frame, highlighting the potential for reusing Gaussians from the previous frame's table.

\begin{figure}[t]
    \includegraphics[width=\linewidth]{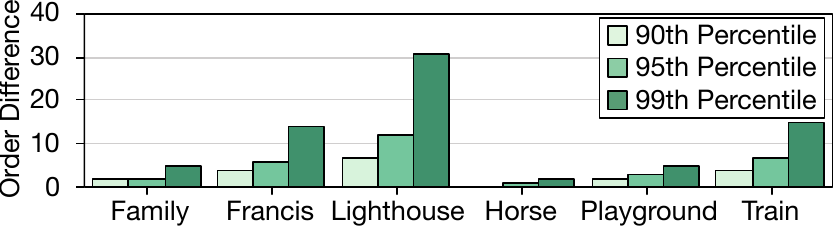}
    \centering
    \caption{Temporal similarity of sort order per tile.}
    \label{fig:opportunity-similarity-sort-order}
    \vspace{-\baselineskip}
\end{figure}

In a second experiment, we measure how the ordering of Gaussians within each tile changes between consecutive frames. 
Figure~\ref{fig:opportunity-similarity-sort-order} illustrates the sorting order differences at the 90th, 95th, and 99th percentiles. 
Results show that 99\% of the sorting order remains largely consistent across consecutive frames. 
Notably, at the 99th percentile, the Gaussian with the greatest shift moves only 31 positions from its original location, a negligible deviation given that each tile contains thousands of Gaussians. 
Extending from the Gaussian retention observed in Figure~\ref{fig:opportunity-similarity-assigned-gaussian}, this result further substantiates the opportunity to leverage temporal similarity between frames to mitigate redundant updates in the Gaussian table.
\newline

\textit{
These findings indicate the need for a bandwidth-optimized sorting technique for on-device 3DGS systems. 
To address this, we propose~\sysname{}, a hardware-algorithm co-designed solution that enables efficient 3DGS rendering through memory-efficient sorting acceleration, leveraging temporal similarity in moving camera scenarios within AR/VR applications. 
}

\section{\sysname{}'s Reuse-and-Update Sorting}
This section introduces the software component of our~\sysname{} solution, designed to efficiently leverage temporal similarity between consecutive frames to facilitate high-resolution 3DGS rendering on edge devices.
By intelligently reusing sorting information across frames, we significantly reduce bandwidth requirements and computational overhead.

\subsection{Design Space Exploration and Considerations}
\label{sec:design-choice}
\niparagraph{Design space exploration of sorting reuse methods.}
When leveraging temporal similarity under moving camera, two conventional strategies are periodic sorting and background sorting.
Periodic sorting intermittently recomputes the full sorting order while skipping sorting in intermediate frames, which reduces average latency but introduces occasional spikes.
Moreover, errors accumulate between refresh intervals, leading to gradual degradation in rendering quality.
In contrast, background sorting~\cite{3dgs:webgl} continuously updates sorting results in parallel with rendering, and each frame uses the most recently prepared results.
This approach mitigates latency spikes but introduces sustained memory traffic, which incurs memory contention and increases average latency.
Moreover, discrepancies between the viewpoints of sorting and rendering frames degrade visual quality.
To address these limitations, we propose an incremental update strategy that reuses the previous frame’s sorting results while applying fine-grained corrections.
Even under abrupt camera motion, this method recovers the correct ordering within a few frames, eliminating the need for full sorting.

\niparagraph{Considerations for incremental update.}
In designing the incremental update strategy, we identify three primary sources of variability that the system must handle:
(1) Change in camera viewpoint can alter Gaussian depths, invalidating the previous ordering.
(2) New Gaussians can become visible in a specific tile.
(3) Gaussians no longer relevant may disappear from the tile.
Handling these factors is critical for robust incremental sorting, ensuring consistent accuracy under dynamic viewpoints and scene content.

\subsection{Flow of Reuse-and-Update Sorting}

\begin{figure}[t]
    \centering
    \includegraphics[width=\linewidth]{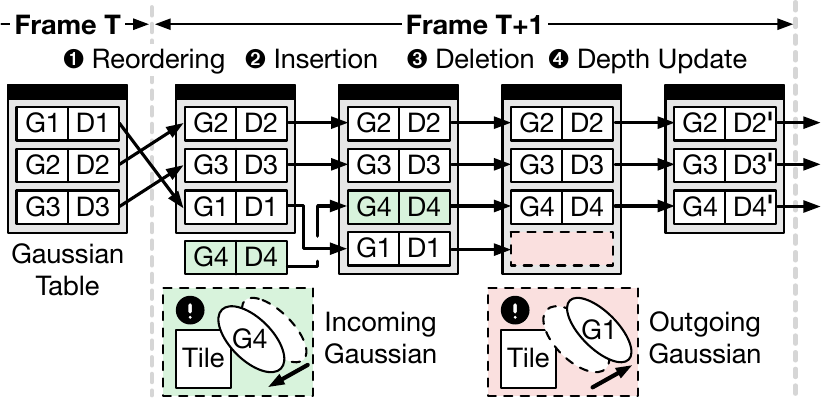}
    \caption{Overview of reuse-and-update sorting.}
    \label{fig:algorithm-workflow}
    \vspace{-\baselineskip}
\end{figure}
Figure~\ref{fig:algorithm-workflow} shows the flow of the reuse-and-update sorting algorithm, consisting of four main operations.

\niparagraph{\circleN{1} Reordering.}
We reuse the Gaussian table from the previous frame.
However, when the viewpoint changes, the depth order of Gaussians changes, requiring reordering.
To handle this efficiently, we employ \emph{Dynamic Partial Sorting}, which performs partial sorting within on-chip memory and minimizes off-chip accesses.
Section~\ref{sec:algorithm-adaptive-sorting} provides further details.

\niparagraph{\circleN{2} Insertion.}
We collect newly visible Gaussians entering a tile (i.e., incoming Gaussians) separately and insert them into the Gaussian table.
Since the number of incoming Gaussians is typically small compared to the Gaussian table, this insertion incurs minimal computational overhead.

\niparagraph{\circleN{3} Deletion.}
We identify Gaussians that move out of a tile (i.e., outgoing Gaussians) due to camera motion and remove them from the Gaussian table at each iteration.

After completing the three steps above, we use the updated Gaussian table for rasterization.
During rasterization, the system fetches necessary Gaussian features (e.g., position, depth, and color) to compute pixel values of the scene.

\niparagraph{\circleN{4} Depth Update.}
During rasterization, we update the depth values in the sorted Gaussian table.
We exploit the availability of these values in this stage to enable on-the-fly depth refresh for subsequent frames.
This design eliminates costly irregular accesses that would otherwise be required to fetch depth values.
Section~\ref{sec:depth-update-detail} provides further details.

After completing reordering, insertion, deletion, and depth update, the system passes the Gaussian table with updated depth values to the next frame’s 3DGS rendering pipeline.

\subsection{Reordering the Reused Gaussian Table}
\label{sec:algorithm-adaptive-sorting}

\SetKwInOut{Input}{input}
\SetKwInOut{Output}{output}
\SetKwComment{Comment}{\textsf{//} }{}

\begin{algorithm}[t]
\small
\caption{Dynamic Partial Sorting}
\label{alg:temporal-adaptive-sorting}

\KwIn{$
    \hspace*{1.0em}{I}: \small\textit{Current\enspace Frame\enspace Iteration\enspace Number} \newline
    \hspace*{1.0em}{G_{I-1}}: \small\textit{Previous\enspace Sorted\enspace Gaussian\enspace Table\enspace of\enspace Tile} \newline
    \hspace*{1.0em}{L}: \small\textit{Size\enspace of\enspace Gaussian\enspace Table\enspace of\enspace Tile} \newline
    \hspace*{1.0em}{C}: \small\textit{Size\enspace of\enspace Chunk\enspace for\enspace Chunk\enspace Sorting}
$}

\KwOut{$
    \hspace*{0.15em}{G_I}: \small\textit{Current\enspace Sorted\enspace Gaussian\enspace Table\enspace of\enspace Tile}
$}

\SetKwFunction{Slice}{\textnormal{Slice}}
\SetKwFunction{Sort}{\textnormal{Sort}}


\Comment{\textit{\textsf{\small{Interleaving Sorting Boundaries}}}}

\If{$I\enspace mod\enspace 2 \equiv 1$}{
    $range \gets (start: 0,\enspace end: C)$ \;
}
\Else{
    $range \gets (start: 0,\enspace end: \lfloor{\frac{C}{2}}\rfloor)$ \;
}

\While {true}{ \label{line:while-start}
    \Comment{\textit{\textsf{\small{Chunk-based Partial Sorting}}}}
    $S \gets \Slice(G_{I-1},\enspace range)$ \; \label{line:slice}
    $S^{'} \gets \Sort(S)$ \; \label{line:sort}
    $\Slice(G_I,\enspace range) \gets S^{'}$ \; \label{line:slice-store}
    
    \Comment{\textit{\textsf{\small{Update Sorting Parameters}}}}
    \If{$range.start + C \ge L$}{
        break \; \label{line:break}
    }
    $range.start \gets range.start + C$ \;
    $range.end \gets min(range.end + C,\enspace L)$ \;
} \label{line:while-end}

\end{algorithm}

In order to exploit the Gaussian tables carried over from previous frames, we devise a sorting algorithm that leverages temporal locality to make on-the-fly corrections.
We refer to this algorithm as \emph{Dynamic Partial Sorting}.
Algorithm~\ref{alg:temporal-adaptive-sorting} outlines our \emph{Dynamic Partial Sorting} approach, comprising two main strategies: chunk-based partial sorting and interleaving the sorting boundaries.
By leveraging the sorted table from the previous frame, our method avoids a full global sort each time, reducing off-chip bandwidth usage while preserving accurate ordering over consecutive frames.

\niparagraph{Chunk-based partial sorting (lines 5--12)}
We observe that the Gaussian sorting order within each tile remains largely consistent across adjacent frames.
Based on this observation, our design partitions the Gaussian table into small chunks that fit within on-chip memory, and sorts each chunk independently.
In our implementation, each chunk stores up to 256 Gaussians in on-chip memory.
During each iteration, we read one chunk (line~\ref{line:slice}) from DRAM into on-chip memory, sort it in place (line~\ref{line:sort}), and write back the results (line~\ref{line:slice-store}).
This local sorting approach significantly reduces bandwidth usage by retrieving and writing back each chunk only once, unlike conventional global sorting methods that make multiple passes over the entire table, repeatedly scanning and rewriting it.
The algorithm continues chunk by chunk until the full table has been processed (lines 9--10).

\begin{figure}[t]
    \centering
    \includegraphics[width=\linewidth]{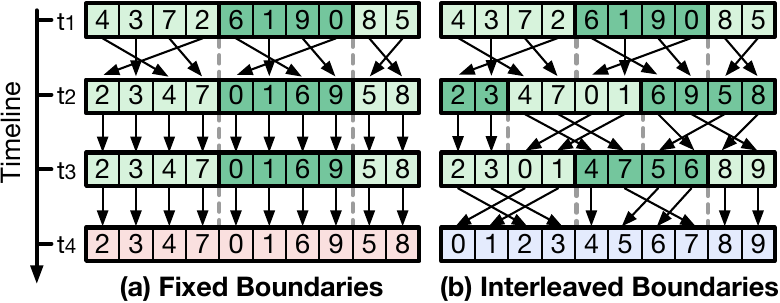}
    \caption{Comparison of two different partial sorting.}
    \label{fig:algorithm-odd-even-different-range-strategy}
    \vspace{-\baselineskip}
\end{figure}

\niparagraph{Interleaving sorting boundaries (lines 1--4).}
Relying on a static partition can lead to inaccuracy if Gaussians need to cross chunk boundaries.
Figure~\ref{fig:algorithm-odd-even-different-range-strategy}(a) illustrates the limitation of performing local sorting.
Suppose at time $t_1$, the depths of Gaussians change due to camera viewpoint shifts, and requires reordering to depths 0 to 9.
Because we only apply partial sorting within fixed chunks, Gaussians cannot cross these boundaries, even after multiple sorting iterations at subsequent times $t_2$, $t_3$, and $t_4$.
In contrast, Figure~\ref{fig:algorithm-odd-even-different-range-strategy}(b) demonstrates our strategy to interleave sorting boundaries.
Initially, at time $t_1$, we perform the same local sorting as in the fixed-boundary approach.
However, starting from the next iteration ($t_2$), we interleave sorting boundaries, shifting chunk boundaries by half the chunk size.
This staggered adjustment allows Gaussians to cross previous chunk boundaries, progressively reaching their correct positions.
By repeating this process over subsequent iterations, Gaussians can freely move toward their correct sorting positions.
As illustrated, by time $t_4$, all Gaussians have successfully reached their intended positions, highlighting the effectiveness of our interleaving sorting boundary method.

\niparagraph{Single off-chip sorting pass.}
In our design, we retrieve each chunk from DRAM and write it back only once.
While multiple sorting passes are possible, it introduces a trade-off between accuracy and memory traffic, depending on the number of off-chip passes.
Increasing the number of passes guarantees more accurate Gaussian ordering, which improves rendering quality, but incurs memory traffic proportional to the number of passes.
In practice, we observe that a single sorting pass introduces only negligible accuracy degradation (e.g., less than 0.1~dB).
Because additional passes provide marginal benefit, we adopt a single off-chip sorting pass to minimize memory traffic.

\niparagraph{Accuracy restoration.}
\emph{Dynamic Partial Sorting} may require a few iterations to reestablish accurate ordering, potentially degrading accuracy.
However, applying this technique reduces off-chip accesses and enhances sorting performance, leading to faster rendering.
This creates a positive feedback loop: faster rendering enables more frequent sorting and updates, which, in turn, maintains accurate ordering during continuous camera movement.
As a result, this technique leads to negligible accuracy degradation.

\subsection{Bandwidth-Efficient Depth Update}
\label{sec:depth-update-detail}

\niparagraph{Challenges in per-frame depth refresh.}
To maintain an up-to-date sorted Gaussian table, we must update the depth values within the table.
A naive approach would fetch each Gaussian’s updated depth from the large off-chip feature table after rendering.
However, this approach incurs substantial random DRAM accesses because it performs a per-Gaussian depth refresh of the Gaussian table for every tile.
Since our design prioritizes minimizing off-chip bandwidth, such random accesses significantly degrade performance.

\niparagraph{Deferred depth update.}
We observe that rasterization already fetches each Gaussian’s full feature information from the off-chip table.
Hence, rather than incurring a second memory pass for depth updates, we piggyback on this access by directly overwriting the depth values in the sorted Gaussian table during rasterization, eliminating redundant memory operations.
However, this design defers the depth update such that, for any given frame, sorting relies on depth values that are one frame stale.
In practice, this one-frame delay introduces only negligible ordering error without degrading rendering quality.
Furthermore, without this optimization, depth updates incur an additional memory access, doubling the memory traffic per tile.
As a result,\sysname{} without this strategy exhibits 33.2\% higher memory traffic compared to the full~\sysname{} design.
Likewise, this optimization significantly reduces random off-chip memory accesses.

\begin{figure}[t]
    \centering
    \includegraphics[width=\linewidth]{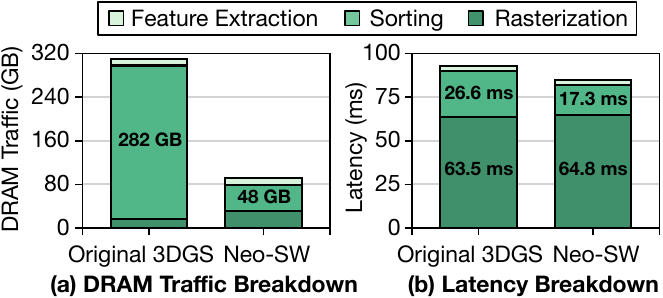}
    \caption{
        Performance comparison between the original 3DGS and~\sysname{}-SW on Orin AGX, with DRAM traffic measured over 60 rendered frames.
    }
    \label{fig:algorithm-comparison}
    \vspace{-\baselineskip}
\end{figure}

\subsection{Performance Implication from~\sysname{} Algorithm}
\label{sec:algorithm-comparison}

\noindent
To evaluate the performance gains of a software-only version of~\sysname{}, we implement a custom CUDA kernel for sorting and modify the rasterization kernel.
We enable \textit{Dynamic Partial Sorting} in the \circleN{1} reordering step by modifying existing sorting libraries from Meta~\cite{meta-cuda} and NVIDIA~\cite{nvidia-cccl}.
We integrate the \circleN{2} insertion and \circleN{3} deletion steps into a merge-sorting process that combines incoming Gaussians with the reused Gaussian table.
Finally, we implement \circleN{4} deferred depth updates during rasterization.
We evaluate this implementation on the NVIDIA Orin AGX platform.

\niparagraph{Limitation of software-only solution.}
Figure~\ref{fig:algorithm-comparison} presents the latency and DRAM traffic of the software-only implementation of~\sysname{}, with DRAM traffic measured over 60 rendered frames.
While the algorithm significantly reduces memory traffic, with 70.4\% overall and 82.8\% during the sorting stage as shown in Figure~\ref{fig:algorithm-comparison}(a), its latency improvement remains modest at only 1.1$\times$, as shown in Figure~\ref{fig:algorithm-comparison}(b).
This modest speedup results from two primary factors.
First, the insertion and deletion operations in the sorting stage induce irregular memory access patterns, degrading spatial locality and limiting SIMD utilization.
Consequently, despite the significant reduction in memory traffic, the sorting stage achieves only a 1.54$\times$ speedup.
Second, as prior work~\cite{lee2024gscore, ye2025gaussian, lee2025vr} highlights, rasterization remains the dominant bottleneck in GPU-based execution, accounting for 68.8\% of total runtime.
Since~\sysname{} specifically targets the sorting stage under the assumption that rasterization has already been accelerated, it delivers inherently limited end-to-end impact on GPU performance.
These inefficiencies underscore the fundamental limitations of GPU-based execution.
Addressing them requires a hardware-software co-designed architecture, which we introduce in the next section.

\section{\sysname{} Accelerator Architecture}
This section presents the hardware architecture of~\sysname{}, which accelerates the 3DGS pipeline with reuse-and-update sorting.
We first provide an overview of the data flow across the engines, followed by detailed descriptions of each engine.

\subsection{Architecture Overview}
\begin{figure}[t]
    \centering
    \includegraphics[width=\linewidth]{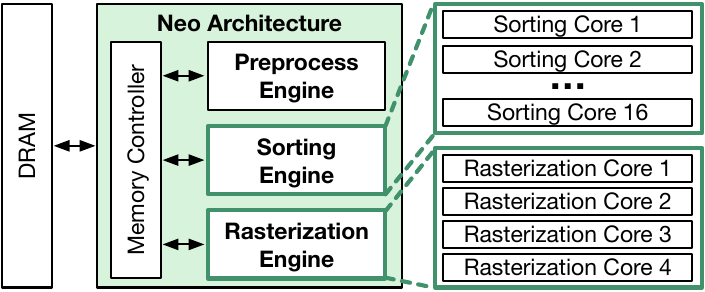}
    \caption{Overall of architecture of \sysname.}
    \label{fig:architecture-overview}
\end{figure}

Figure~\ref{fig:architecture-overview} illustrates the architecture of~\sysname{}.
Our design consists of three main engines, Preprocessing Engine, Sorting Engine, and Rasterization Engine.

\niparagraph{Preprocessing Engine.}
The Preprocessing Engine handles the first two stages of the 3DGS pipeline, namely frustum culling and feature extraction.
For each Gaussian, it determines the tiles where the Gaussian has become newly visible and collects the per-Gaussian information required for rendering.
These operations produce two types of tables:
(1) a feature table, which stores Gaussian attributes required for rasterization, and
(2) incoming Gaussian tables, which record the newly visible Gaussians for each tile.

\niparagraph{Sorting Engine.}
Next, the Sorting Engine performs two types of sorting.
First, it performs \emph{Dynamic Partial Sorting} on the Gaussian tables from the previous frame that contain newly updated depth values.
Second, it sorts the incoming Gaussian tables from the current frame provided by the Preprocessing Engine.
To accelerate both reordering and sorting, the engine employs specialized parallel sorting units.

\niparagraph{Rasterization Engine.}
Finally, the Rasterization Engine performs the last stage of the pipeline, rasterization, using both the sorted Gaussian tables and the feature table.
To eliminate redundant computations, our design adopts subtiling~\cite{lee2024gscore}, which applies $\alpha$-blending only to each subtile for the Gaussians intersecting it.
The Rasterization Engine integrates dedicated hardware for on-the-fly subtiling-based rasterization and for deferred depth updates, which enable reuse-and-update sorting in subsequent frames.

\subsection{Preprocessing Engine}
The Preprocessing Engine consists of projection units and color calculation units for frustum culling and feature extraction, and incorporates duplication units to support the reuse-and-update sorting scheme.

\niparagraph{Conventional preprocessing.}
The projection unit projects Gaussians onto the image plane and performs frustum culling to discard those outside the camera frustum.
Next, the color calculation unit derives view-dependent color for each Gaussian using spherical harmonics.
Together, these units generate a feature table that stores essential rasterization attributes, including color, mean, covariance matrix, opacity, and radius.

\niparagraph{Processing incoming Gaussians.}
The duplication unit uses 2D Gaussian information to identify the tiles intersected by each Gaussian and generates the corresponding Gaussian tables.
Our design adds a verification step that checks whether each Gaussian exists in the previous frame’s Gaussian table.
This step produces the incoming Gaussian tables by allowing the system to process only newly visible Gaussians, thereby enabling the reuse-and-update sorting scheme.
As a result, the unit outputs per-tile Gaussian tables containing the IDs and depth values of newly incoming Gaussians.

\begin{figure}[t]
    \centering
    \includegraphics[width=\linewidth]{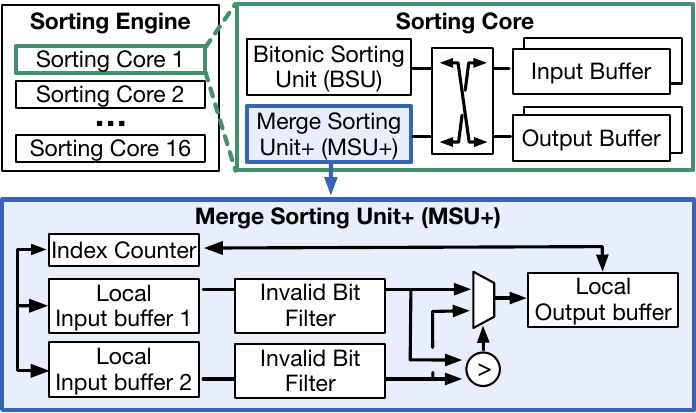}
    \caption{Microarchitecture of Sorting Engine.}
    \label{fig:sorting-engine-architecture}
    \vspace{-\baselineskip}
\end{figure}

\subsection{Sorting Engine}
Figure~\ref{fig:sorting-engine-architecture} shows the microarchitecture of our Sorting Engine, which comprises 16 parallel Sorting Cores. 
Each Sorting Core has a Bitonic Sorting Unit (BSU), a Merge Sorting Unit+ (MSU+), and I/O buffers.
Both input and output buffers employ double buffering to mask memory-access latency. 

\niparagraph{Conventional sorting.}
To perform conventional sorting of the Gaussian table from scratch, the Sorting Core follows standard merge-sort steps.
Each core loads a 256-entry chunk into its input buffer and partitions it into smaller 16-entry sub-chunks that the BSU can process.
The BSU sorts each sub-chunk, and the MSU+ merges the partially sorted results.
The system then writes each fully sorted chunk back to DRAM, processes the remaining chunks, and finally performs a global merge across all sorted chunks.

\niparagraph{Dynamic Partial Sorting.}
Beyond conventional sorting, the Sorting Engine supports \emph{Dynamic Partial Sorting} to reuse the Gaussian table from the previous frame.
Specifically, the engine loads a 256-entry chunk from the previous frame’s table and reorders it in the same manner as conventional sorting, using the BSU to sort sub-chunks followed by an MSU+ merge.
Because no additional merges are required across chunks, the design avoids extra off-chip memory traffic.

\niparagraph{Inserting and deleting entries.}
Alongside conventional sorting and \emph{Dynamic Partial Sorting}, the MSU+ merges the sorted Gaussian table with the incoming Gaussian table and removes outgoing Gaussians from the table.
Specifically, the Preprocessing Engine first generates the incoming Gaussian tables, and the Sorting Engine sorts them using a conventional algorithm before merging them with the sorted Gaussian table from the previous frame.
At the same time, the MSU+ removes outgoing Gaussians based on their valid bits, which were marked as valid or invalid during the previous frame’s rasterization.
This design stems from the observation that, although outgoing Gaussians can be excluded during rasterization using their valid-bit flag, immediately removing them would require costly shifting of subsequent entries.
By deferring memory realignment to the merge step, the MSU+ efficiently deletes invalid entries without incurring the cost of shifting the entries.
Furthermore, it inserts new entries simultaneously, thereby improving performance.

\subsection{Rasterization Engine}
\begin{figure}[t]
    \centering
    \includegraphics[width=\linewidth]{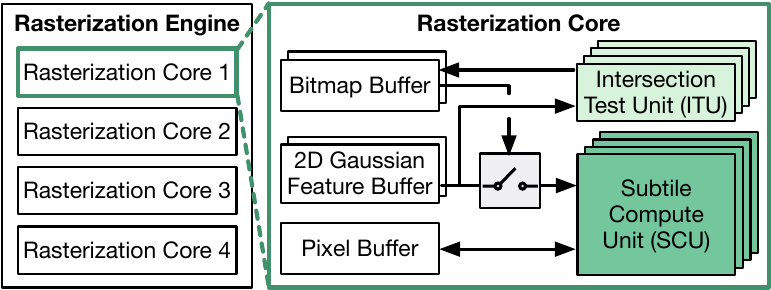}
    \caption{Microarchitecture of Rasterization Engine.}
    \label{fig:Rasterization-core-architecture}
\end{figure}

Figure~\ref{fig:Rasterization-core-architecture} illustrates the microarchitecture of our Rasterization Engine, comprising four Rasterization Cores. 
Each core contains four Subtile Compute Units (SCU), four Intersection Test Units (ITU), and dedicated buffers for bitmaps, 2D Gaussian features, and pixel data. 

\niparagraph{Intersection Test Unit~(ITU).}
Our design adopts subtile-based rasterization, inspired by GSCore~\cite{lee2024gscore}, which subdivides each tile into multiple smaller subtiles.
Since a Gaussian typically intersects only a subset of subtiles within a tile, this approach reduces redundant computations.
To track Gaussian–subtile intersections, GSCore maintains lightweight subtile metadata in the form of a bitmap that indicates whether a Gaussian overlaps each subtile.
However, 3DGS generates these bitmaps early in the pipeline and propagates them through all stages, even though they are only used during rasterization, causing unnecessary memory traffic.
To address this inefficiency, we integrate Intersection Test Units (ITUs) within the Rasterization Core to generate the required bitmaps on the fly.
Specifically, each ITU uses the 2D parameters of a Gaussian to test intersection boundaries and store the resulting bitmap in a local buffer.
Moreover, the ITU plays a key role in reuse-and-update sorting by detecting outgoing Gaussians through a cumulative OR operation.
This operation accumulates intersection bitmaps across all subtiles to flag Gaussians with at least one intersection.
By flipping this bit, the system identifies Gaussians with no intersections and eliminates them in the next sorting stage.

\begin{figure}[t]
    \centering
    \includegraphics[width=\linewidth]{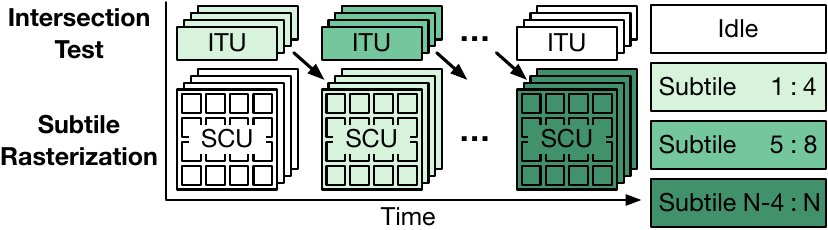}
    \caption{Execution timeline of Rasterization Engine. Intersection Test Unit (ITU) and Subtile Computation Unit (SCU) process subtiles in a pipelined manner.}
    \label{fig:intersection-test-hiding-architecture}
    \vspace{-\baselineskip}
\end{figure}

\niparagraph{Subtile Computation Unit~(SCU).}
Following the intersection tests, each SCU performs $\alpha$-blending to compute pixel values for its assigned subtile.
The system filters Gaussians using the bitmaps generated by the ITUs and routes them only to SCUs whose subtiles intersect with the Gaussian.
After rasterizing the 2D Gaussian feature buffer for a given group of subtiles, the system writes the intermediate pixel values to the pixel buffer and proceeds to the next group.
To maximize efficiency, we pipeline intersection testing with rasterization.
Figure~\ref{fig:intersection-test-hiding-architecture} illustrates this pipelining scheme.
While rasterization for the first group of four subtiles (1–4) must wait for its intersection tests to complete, subsequent subtiles benefit from overlapping execution as the ITUs process the next group of subtiles concurrently with the ongoing rasterization of the current one.
This overlap effectively hides the latency of on-the-fly bitmap generation.

\begin{figure*}[t]
    \centering
    \includegraphics[width=\linewidth]{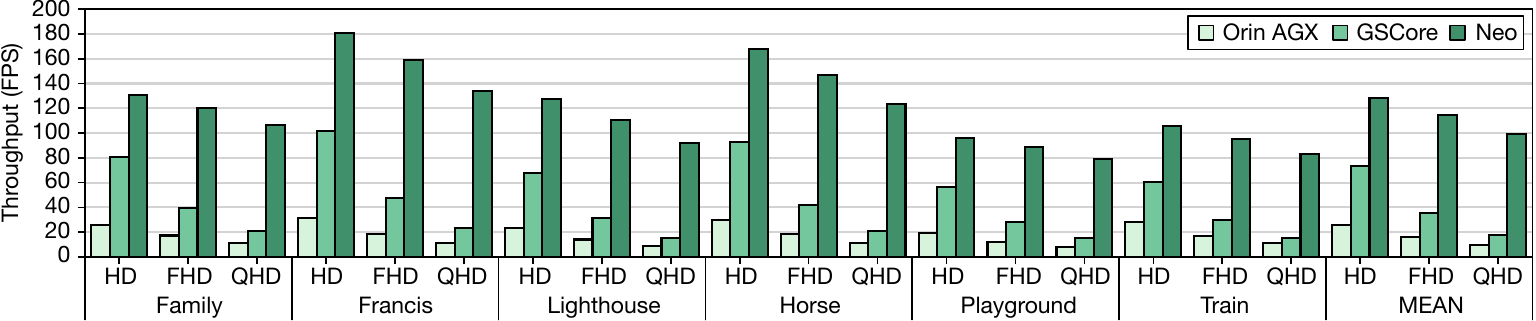}
    \caption{End-to-end system throughput of GSCore and~\sysname{} on six 3D scenes.}
    \label{fig:evaluation-performance-throughput}
\end{figure*}

\niparagraph{Updating table for next frame.} 
After rasterizing all subtiles, the system extracts depth values from the 2D Gaussian feature buffer and retrieves valid bits from the bitmap buffer.
It then updates the corresponding entries in the Gaussian table and forwards the updated table to the next frame.
\begin{table}[t]
    \small
    \caption{Configuration of \sysname system.}
    \vspace{-2ex}
    \label{tab:hardware-configuration}
    \centering
    \begin{tabular}{cl|c}
    \hlinewd{1.0pt}
    \multicolumn{2}{c}{\textbf{Hardware Component}} & \textbf{Configuration} \\
    \hlinewd{1.0pt}
    
    \textbf{\sysname} & Tile Size & 64$\times$64 px \\    
    \hline
    
    \multirow{3}{*}{\shortstack{\textbf{Preprocessing}\\\textbf{Engine}}} & Projection Unit & 4 units \\
    & Color Unit & 4 units\\
    & Duplication Unit & 4 units \\
    \hline

    \multirow{3}{*}{\shortstack{\textbf{Sorting}\\\textbf{Engine}}} & Bitonic Sort Unit & 16 units \\
    & Merge Sort Unit+ & 16 units \\
    & I/O Buffer Size & 64 KB \\
    \hline
    
    \multirow{4}{*}{\shortstack{\textbf{Rasterization}\\\textbf{Engine}}} & Subtile Compute Unit & 16 units \\
    & Intersection Test Unit & 16 units \\
    & Buffer Size & 200 KB \\
    & Subtile Size & 8$\times$8 px \\
    
    \hlinewd{1.0pt}
    \end{tabular}
\end{table}

\section{Evaluation}

\subsection{Methodology}
\label{sec:evaluation-methodology}

\niparagraph{Benchmarks.}
We select six scenes from the Tanks and Temples dataset~\cite{tankandtemples:tog:2017}, namely Family, Francis, Horse, Lighthouse, Playground, and Train, as representative benchmarks for evaluating 3DGS rendering quality.
We capture each frame at 30 FPS in UHD resolution (3840$\times$2160).
Following the standard training procedure, we use 400 images per scene and train for 200K iterations to ensure model convergence.
For inference, we configure each scene to one of three target resolutions: HD (1280$\times$720), FHD (1920$\times$1080), or QHD (2560$\times$1440), and evaluate how many unique frames the system renders per second, with each frame using a distinct camera pose from the original sequence of the dataset.

\niparagraph{Hardware development and synthesis.}
Table~\ref{tab:hardware-configuration} shows the configurations of our hardware.
We prototype~\sysname{} architecture at RTL level using Verilog, and measure power, area, and timing parameters using Synopsys Design Compiler with ASAP7~\cite{asap7} 7~nm library.
We measure the power and area of on-chip buffers using CACTI~\cite{cacti} under 22~nm technology and scale them to 7~nm using DeepScaleTool~\cite{deepscaletool}.

\niparagraph{Baseline.}
We evaluate~\sysname against two baseline systems: NVIDIA Orin AGX 64GB (Orin AGX), GSCore.
Orin AGX~\cite{karumbunathan2022nvidia} is a high-performance on-device platform designed for autonomous systems.
It supports up to 60W of power and provides 204.8~GB/s of memory bandwidth.
This baseline enables evaluation of rendering performance on real on-device hardware.
GSCore~\cite{lee2024gscore} originally features four sorting and rasterization cores.
To support high-resolution workloads and enable a fair comparison with~\sysname{}, which includes 16 hardware units, we scale GSCore to 16 cores.

\niparagraph{Cycle accurate simulator.}
We measure the latency of GSCore, and~\sysname using a cycle-accurate simulator based on the timing parameters obtained from RTL synthesis, with off-chip memory modeled as LPDDR4 based on Ramulator~\cite{kim2015ramulator}.

\begin{figure}[t]
    \centering
    \includegraphics[width=\linewidth]{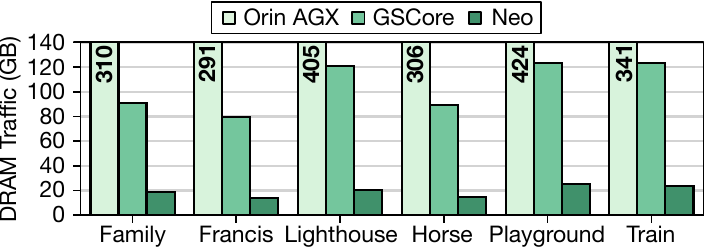}
    \caption{Comparison of DRAM traffic (GB) between~\sysname{} and GSCore for rendering 60 frames.}
    \label{fig:evaluation-performance-traffic}
    \vspace{-\baselineskip}
\end{figure}

\subsection{Performance Results}
\label{sec:evaluation-performance-results}

\niparagraph{End-to-end throughput.}
Figure~\ref{fig:evaluation-performance-throughput} presents the end-to-end rendering throughput for six scenes rendered at three target resolutions (HD, FHD, and QHD) on Orin AGX, GSCore, and~\sysname{}.
Across all scenes and resolutions,~\sysname{} consistently outperforms both Orin AGX and GSCore, achieving average 5.0$\times$, 7.2$\times$, and 10.0$\times$ speedup over Orin AGX and 1.8$\times$, 3.3$\times$, and 5.6$\times$ over GSCore for HD, FHD, and QHD, respectively.
The performance gains are particularly significant at higher resolutions, where sorting bottlenecks intensify, demonstrating~\sysname{}’s effectiveness in addressing bandwidth bottlenecks through temporal similarity.
Unlike Orin AGX and GSCore, which sort from scratch in every frame,~\sysname{} reduces bandwidth usage with a reuse-and-update sorting approach.
It applies \emph{Dynamic Partial Sorting} to tables from the previous frame while performing conventional sorting only on small incoming Gaussian tables.
Notably,~\sysname{} achieves an average throughput of 99.3~FPS at QHD resolution, meeting the real-time rendering requirement.

\niparagraph{End-to-end memory traffic.}
Figure~\ref{fig:evaluation-performance-traffic} shows the required DRAM traffic to render 60 frames in QHD resolution on Orin AGX, GSCore, and~\sysname{}.
For Orin AGX, rendering 60 frames requires an average of 346.5~GB across six scenes, compared to 104.6~GB for GSCore and 19.6~GB for~\sysname{}, representing a 94.4\% and 81.3\% reduction over OrinAGX and GSCore, respectively.
This reduction is largely due to significant traffic savings achieved through reuse-and-update sorting.
As a result, even in scenarios with limited bandwidth (e.g., 51.2~GB/s), the system can perform computations without being bottlenecked by the bandwidth constraints.

\begin{table}[t]
    \small
    \caption{Quality comparison of original 3DGS and~\sysname.}
    \vspace{-2ex}
    \label{tab:evalution-accuracy}
    \centering
    \begin{tabular}{c|cc|cc}
    \hlinewd{1.0pt}
    \multirow{2}{*}{\textbf{Scene}} & \multicolumn{2}{c}{\textbf{Original 3DGS}} \vline & \multicolumn{2}{c}{\textbf{~\sysname}} \\
    \cline{2-5}
    \noalign{\vskip 1pt}
    & \textbf{PSNR$\uparrow$} & \textbf{LPIPS$\downarrow$} & \textbf{PSNR$\uparrow$} & \textbf{LPIPS$\downarrow$} \\
    \hlinewd{1.0pt}
    Family & 28.2 & 0.096 & 28.1~\textcolor{red}{($\blacktriangledown$0.1)} & 0.097~\textcolor{red}{($\blacktriangle$0.001)} \\
    \hline
    Francis & 28.9 & 0.203 & 28.9~\textcolor{ForestGreen}{($\bullet$)} & 0.203~\textcolor{ForestGreen}{($\bullet$)} \\
    \hline
    Horse & 28.0 & 0.110 & 27.9~\textcolor{red}{($\blacktriangledown$0.1)} & 0.110~\textcolor{ForestGreen}{($\bullet$)} \\
    \hline
    Lighthouse & 26.0 & 0.096 & 26.0~\textcolor{ForestGreen}{($\bullet$)} & 0.096~\textcolor{ForestGreen}{($\bullet$)} \\
    \hline
    Playground & 25.1 & 0.208 & 25.0~\textcolor{red}{($\blacktriangledown$0.1)} & 0.208~\textcolor{ForestGreen}{($\bullet$)} \\
    \hline
    Train & 25.0 & 0.113 & 24.9~\textcolor{red}{($\blacktriangledown$0.1)} & 0.113~\textcolor{ForestGreen}{($\bullet$)} \\
    \hlinewd{1.0pt}
    \end{tabular}
\end{table}
\begin{table}[t]
    \small
    \caption{Evaluated GSCore and~\sysname{} accelerators.}
    \vspace{-2ex}
    \label{tab:area-and-power}
    \centering
    \begin{tabular}{ccccc}
    \hlinewd{1.0pt}
    \multirow{2}{*}{\textbf{Device}} & \multirow{2}{*}{\textbf{Technology}} & \multirow{2}{*}{\textbf{Frequency}} & \textbf{Area} & \textbf{Power} \\
    & & & \textbf{(\textit{mm\textsuperscript{2}})} & \textbf{(\textit{mW})} \\
    \hlinewd{1.0pt}
    \textbf{GSCore} & \multirow{2}{*}{7~nm} & \multirow{2}{*}{1~GHz} & 0.417 & 719.9 \\
    \textbf{\sysname} & & & 0.387 & 797.8 \\
    \hlinewd{1.0pt}
    \end{tabular}
\end{table}

\begin{table}[t]
    \small
    \caption{Area and power breakdown of hardware components in~\sysname{} accelerator.}
    \vspace{-2ex}
    \label{tab:area-and-power-breakdown}
    \centering
    \begin{tabular}{l|cc}
    \hlinewd{1.0pt}
    \noalign{\vskip 2.0pt}
    \multicolumn{1}{c}{\textbf{Component}} & \textbf{Area (\textit{mm\textsuperscript{2}})} & \textbf{Power (\textit{mW})} \\
    \noalign{\vskip 2.0pt}
    \hlinewd{1.0pt}
    
    \textbf{Preprocessing Engine}  & \textbf{0.026} & \textbf{194.9} \\
    \hline
    
    Merge Sort Unit+ & 0.005 & 12.4 \\
    Bitonic Sort Unit & 0.008 & 75.0 \\
    Buffers + others & 0.040 & 71.6 \\
    \textbf{Sorting Engine} & \textbf{0.053} & \textbf{159.0} \\
    \hline

    Subtile Compute Unit & 0.228 & 375.0 \\
    Intersection Test Unit & 0.030 & 58.7 \\
    Buffers + others & 0.050 & 10.2 \\
    \textbf{Rasterization Engine} & \textbf{0.308} & \textbf{443.9}\\
    \hlinewd{1.0pt}

    \textbf{Total} & \textbf{0.387} & \textbf{797.8} \\    
    \hlinewd{1.0pt}
    \end{tabular}
\end{table}



\niparagraph{Rendering quality.}
Table~\ref{tab:evalution-accuracy} compares the rendering quality of GSCore and~\sysname{}.
We evaluate rendering quality using standard graphics metrics: PSNR, where higher values indicate better image fidelity, and LPIPS, where lower values indicate better perceptual quality.
Across all scenes, the maximum observed difference is less than 0.1 dB in PSNR and 0.001 in LPIPS, an imperceptible level of quality degradation~\cite{potamoi,cicero:isca:24}.
These results highlight the effectiveness of~\sysname{}’s reuse-and-update sorting mechanism in maintaining high rendering quality while significantly reducing sorting overhead by exploiting temporal similarity.

\niparagraph{Area and power.}
Table~\ref{tab:area-and-power} compares the evaluated area and power of~\sysname{} and GSCore. 
For a fair comparison, we scale GSCore's area and frequency estimates down to 7~nm using DeepScaleTool~\cite{deepscaletool}, as it was originally synthesized in a 28~nm process. 
The results show that~\sysname{} achieves a slightly smaller total area than GSCore with a marginal increase in power consumption. 
To further understand the area and power overhead, Table~\ref{tab:area-and-power-breakdown} provides a detailed breakdown of our~\sysname{} accelerator, showing that its additional hardware components (Merge Sort Unit+ and Intersection Test Unit) together account for 9.04\% and 8.91\% of the total area and power consumption, respectively. 
With minimal overhead, our additional hardware block delivers high throughput under high-resolution settings.

\begin{figure}[t]
    \centering
    \includegraphics[width=\linewidth]{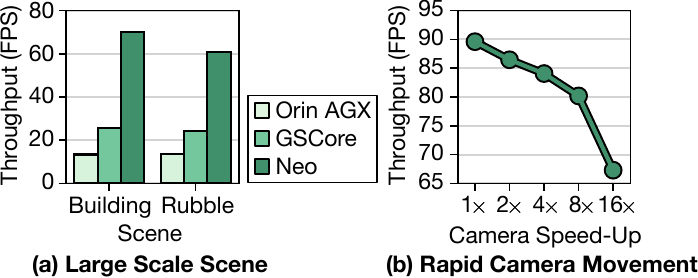}
    \caption{Throughput of 3DGS rendering systems in extreme AR/VR scenarios: (a) Large-Scale Scene, and (b) Rapid Camera Movement.}
    \label{fig:evaluation-varing-constraints}
    \vspace{-\baselineskip}
\end{figure}

\niparagraph{Performance results of extreme AR/VR scenarios.}
Figure~\ref{fig:evaluation-varing-constraints} shows the throughput of~\sysname{} under the more stringent constraints of AR/VR scenarios, including large scale scene rendering and rapid camera movement. 
Figure~\ref{fig:evaluation-varing-constraints}(a) shows the results of Building and Rubble scenes from Mill 19~\cite{turki:2022:mega-nerf} which features high-resolution aerial imagery commonly used to represent complex, large-scale scene.
\sysname{} delivers an average throughput of 65.2~FPS whereas both Orin AGX and GSCore struggle to meet the high frame rate requirements, dropping below 13.6~FPS and 24.9~FPS, respectively. 
Figure~\ref{fig:evaluation-varing-constraints}(b) presents~\sysname{}'s performance under different levels of rapid camera movement (2$\times$, 4$\times$, 8$\times$ and 16$\times$). 
In these scenarios, although Gaussian reusability decreases under rapid camera motion,~\sysname{} maintains a frame rate above 60 FPS, satisfying the conservative SLO requirement for rendering.
These results demonstrate the effectiveness of ~\sysname{}'s reuse-and-update sorting mechanism, exhibiting superior rendering performance across extreme AR/VR scenarios. 

\subsection{Ablation Studies}
\label{sec:ablation-studies}

\begin{figure}[t]
    \centering
    \includegraphics[width=\linewidth]{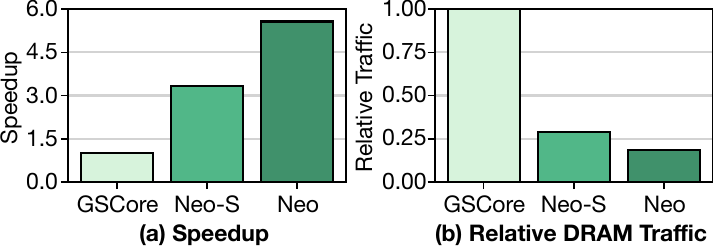}
    \caption{
        Speedup and DRAM traffic normalized to GSCore. ~\sysname{}-S replaces GSCore Sorting Engine with~\sysname{}'s Sorting Engine.
    }
    \label{fig:evaluation-performance-gain-from-hardware}
\end{figure}

\niparagraph{Performance breakdown of~\sysname{} hardware.}
The hardware of~\sysname{} consists of two main components: the Sorting Engine and the Rasterization Engine.
The Sorting Engine implements the first three steps of~\sysname{}’s reuse-and-update sorting algorithm: reordering, insertion, and deletion, while the Rasterization Engine performs subtile-based rasterization and depth update.
Figure~\ref{fig:evaluation-performance-gain-from-hardware} shows the incremental benefits of integrating these components into GSCore.
Although GSCore alone does not support~\sysname{}’s algorithm, adding the Sorting Engine (Neo-S) enables reuse-and-update sorting, reducing memory traffic by 71.1\% and improving performance by 3.3$\times$.
However, without hardware support for depth update, the system requires separate post-processing to update Gaussian table metadata (e.g., depth, valid bit), which incurs additional delay.
Integrating the Rasterization Engine removes this overhead, achieving a further 35.8\% traffic reduction and an additional 1.7$\times$ speedup.
While the Sorting Engine provides substantial benefits, full algorithm support requires a co-designed sorting and rasterization engine.
Accordingly,~\sysname{} adopts a holistic hardware design that integrates both components to maximize performance.

\niparagraph{Comparison with existing sorting methods.}
In Section~\ref{sec:design-choice}, we explore the design space of sorting reuse methods, including periodic sorting, background sorting, and incremental update sorting.
Based on this analysis, we adopt incremental update sorting in our design.
To evaluate its effectiveness in terms of latency and rendering quality, we compare it against two alternative strategies on the~\sysname{} hardware: (1) periodic sorting and (2) background sorting.
We also consider (3) hierarchical sorting, originally proposed in GSCore, which accelerates sorting by combining coarse-grained and fine-grained sorting.
We evaluate hierarchical sorting in conjunction with incremental update sorting to quantify the benefits of the \emph{Dynamic Partial Sorting}.

\begin{figure}[t]
    \centering
    \includegraphics[width=\linewidth]{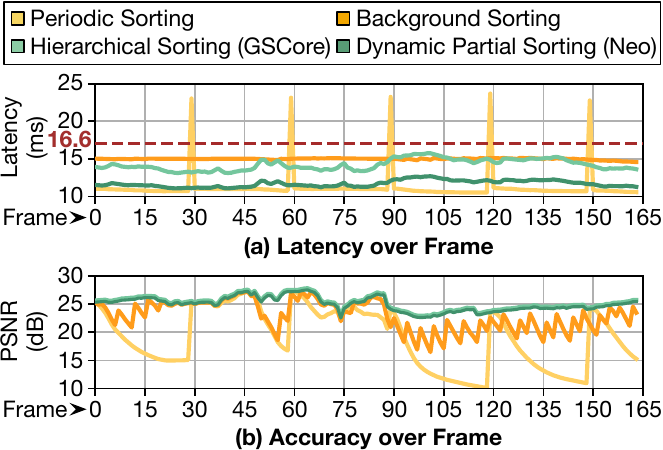}
    \caption{
        Latency and rendering quality across frames for four sorting reuse methods.
    }
    \label{fig:evaluation-design-choice-comparison}
    \vspace{-\baselineskip}
\end{figure}

Figure~\ref{fig:evaluation-design-choice-comparison} compares the three sorting methods.
(1) Periodic sorting achieves lower average latency than~\sysname{} by avoiding continuous updates.
However, it introduces periodic latency spikes that violate the 16.6~ms SLO for 60 FPS and suffers from severe quality degradation due to error accumulation between updates.
(2) Background sorting maintains relatively stable latency by continuously sorting in the background.
Despite this, it incurs higher average latency than~\sysname{} and degrades rendering quality due to temporal viewpoint discrepancies between sorting and rendering.
(3) Hierarchical sorting accurately sorts reused Gaussians and delivers rendering quality comparable to~\sysname{}.
However, it requires multiple passes over off-chip memory, which increases latency.
These results demonstrate the effectiveness of~\sysname{}’s sorting scheme and support our design choice.

\section{Related Work}
\label{sec:related_work}

\niparagraph{View synthesis acceleration.}
Neural rendering has driven significant advances, prompting the architecture community to explore hardware acceleration~\cite{li2025uni, gennerf:isca:23, cicero:isca:24, srender:micro:24, neurex:isca:23, ngpc:isca:23, instant3d:isca:23, fusion3d:micro:24, cambricon-r:mirco:23, lee202466, liu2025cambricon, pei2025gcc}.
With the emergence of 3D Gaussian Splatting (3DGS), research focus has shifted accordingly.
GSCore~\cite{lee2024gscore} introduces hierarchical sorting and an optimized rasterization pipeline.
GBU~\cite{ye2025gaussian} improves efficiency by reusing rasterization results.
VR-Pipe~\cite{lee2025vr} repurposes the GPU rasterization engine with microarchitectural optimizations.
MetaSapiens~\cite{metasapiens:asplos:25} applies load balancing techniques to tile-based rasterization.
For training, ARC~\cite{durvasula2025arc} and GSArch~\cite{he2025gsarch} mitigate atomic overheads using warp-level reductions and gradient filtering.
GauSPU~\cite{gauspu:micro:24} targets SLAM-integrated 3DGS with a sparsity-adaptive rasterizer and a relaxed-memory backpropagation engine.
Our work builds on prior 3DGS accelerators that address rasterization bottlenecks and shifts the focus to a more practical constraint in high resolution rendering, revealing sorting as the next critical performance limiter.
~\sysname{} targets this bottleneck with a lightweight, reuse-aware sorting engine that complements prior efforts.

\niparagraph{Memory-efficient 3DGS.}
Another branch of research reduces the memory footprint of 3DGS through pruning and quantization.
Pruning eliminates low-impact Gaussians based on importance metrics such as opacity~\cite{lightgaussian:nips:25, compgs:eccv:24} or $\alpha$-blending weights~\cite{metasapiens:asplos:25, fang2024mini}.
Quantization techniques, including vector quantization~\cite{lightgaussian:nips:25} and learned compression~\cite{fang2024mini, papantonakis:i3d:24, eagles:eccv:24, compgs:eccv:24}, reduce both memory and storage overhead.
However, these approaches require retraining or fine-tuning.
In contrast,~\sysname{} introduces an orthogonal sorting scheme that requires no retraining.
Moreover, it complements existing methods, enabling further gains in bandwidth efficiency.

\niparagraph{Leveraging temporal similarity.}
Numerous studies on real-time streaming systems~\cite{dutson2023eventful, song2024cmc, buckler2018eva2, song2020vr, parger2022deltacnn, kim2024dacapo, li2020reducto, taranco2023deltalta, hwang2022cova, ying2022exploiting, zhao2020deja, meng2020coterie, feng2019asv, hwang2025dejavu, zhu2018euphrates} have been proposed to reduce computation while maintaining accuracy.
Across diverse domains, one of the most effective strategies is to exploit temporal redundancy.
In graphics, prior work~\cite{glasbey1998review, schollmeyer2017efficient, li2019deltavr, chen1995quicktime, martin2019synthesising, cicero:isca:24} applies this principle through pixel-wise reuse, known as warping.
Neural rendering also benefits from temporal redundancy.
For example, Potamoi~\cite{potamoi} leverages it to bypass MLP overheads in NeRF through pixel reuse.
However, 3DGS performs Gaussian-wise feature extraction and tile-wise sorting and rasterization, which limit the applicability of pixel-wise reuse and fail to address the memory traffic induced by sorting.
In contrast, Lumina~\cite{feng2025lumina} leverages temporal similarity at the sorting stage but performs background sorting, which continuously consumes memory bandwidth, incurs contention, and increases average latency, as discussed in Section~\ref{sec:design-choice} and Section~\ref{sec:ablation-studies}.
~\sysname{} instead exploits temporal similarity at the sorting stage through an incremental update strategy, effectively reducing the associated memory traffic.
\section{Conclusion}

Real-time on-device 3D Gaussian Splatting (3DGS) rendering demands both low latency and high frame generation rate, yet existing solutions struggle to meet these conflicting requirements under stringent resource constraints.
This work identifies Gaussian sorting as a key bottleneck in the 3DGS inference pipeline, and presents~\sysname{}, an on-device acceleration solution that introduces a reuse-and-update sorting algorithm and a hardware-accelerated sorting pipeline to reduce redundant computations and alleviate memory bandwidth pressure.
By tackling this challenge,~\sysname{} takes a step toward enabling real-time, on-device generative virtual worlds, bringing us closer to immersive AR/VR experiences.

\bibliographystyle{ACM-Reference-Format}
\bibliography{reference}

\end{document}